\documentclass[11pt]{article}
\usepackage{geometry}
\usepackage[T1]{fontenc}
\usepackage{graphicx}
\usepackage{amsmath}
\usepackage{booktabs}
\usepackage{multirow}
\usepackage{makecell}
\usepackage{tabularx}
\usepackage{listings}
\usepackage{fvextra}
\usepackage{placeins}
\usepackage{algorithm}
\usepackage{algpseudocode}
\usepackage[numbers]{natbib}
\usepackage{url}
\usepackage{hyperref}
\usepackage{orcidlink}
\usepackage{amssymb}

\begin{document}

\title{A Methodology to Assess Power Modeling in Energy-Aware Federated Learning on Heterogeneous Mobile Devices}

\author{
Chaimae Jallouli$^1$ \orcidlink{0009-0007-0656-6573} \and
Karim Boubouh$^2$ \orcidlink{0000-0002-0920-2165} \and
Robert Basmadjian$^1$ \orcidlink{0000-0003-2951-4243}
}

\maketitle
\begin{center}
\small
$^1$Mohammed VI Polytechnic University, Benguerir, Morocco\\
$^2$Khalifa University, Abu Dhabi, UAE
\end{center}

\footnotetext{This version of the contribution has been accepted 
for publication, after peer review, but is not the Version of Record 
and does not reflect post-acceptance improvements, or any corrections. 
The Version of Record will be available online at: 
\url{https://doi.org/[DOI~to~appear]} once published in the proceedings 
of Networked Systems (NETYS 2026), Springer Nature. Use of this Accepted 
Version is subject to the publisher's Accepted Manuscript terms of use: 
\url{https://www.springernature.com/gp/open-research/policies/accepted-manuscript-terms}}

\begin{abstract}

Estimating CPU power on heterogeneous ARM-based commodity devices is challenging due to limited access to CPU's voltage domains. As a result, state-of-the-art energy-aware Federated Learning (FL) frameworks typically rely on simplified approximate power models to estimate computation energy, rather than the more accurate analytical CMOS-based model. To bridge this gap, we propose a reproducible CPU power estimation methodology combined with a rail-to-cluster mapping technique to retrieve cluster-level supply voltage. We evaluate our approach on two commodity Android devices and show that the analytical model predicts CPU power with errors below 10\%, whereas the approximate model incurs errors of up to 959\%. Using AnycostFL, a state-of-the-art energy-aware FL framework, we show that the analytical model achieves the same 80\% model accuracy while consuming 1.4× less energy than the approximate model. These results highlight that approximate models can severely misestimate computation energy and lead to suboptimal decisions. This work facilitates the use of analytical CPU power models on heterogeneous multi-cluster ARM-based mobile SoCs without additional hardware support or external power measurement tools. 

\end{abstract} 
\section{Introduction}

Green computing has shifted toward energy-aware solutions \cite{murugesan2008harnessing}. As ARM-based devices gain compute power, they are replacing x86 architectures due to superior performance-per-watt \cite{gupta2021changing} and are increasingly hosting on-device Machine Learning (ML) tasks \cite{dhar2021survey}. Unlike x86 processors, which offer OS-accessible metrics like RAPL \cite{raffin2024dissecting}, ARM processors typically only expose frequency. This limited observability hinders analytical CMOS-based power models common for x86 \cite{becker2017software}, with direct implications for battery-powered devices where maintaining performance under tight energy budgets requires accurate power estimation \cite{walker2016accurate}. Since on-device ML is CPU-dominated \cite{shi2022toward}, modeling inaccuracies directly impact energy-aware optimization and the reliability of on-device frameworks.

At the same time, a new generation of on-device ML appeared \cite{basmadjian2022advantages}, such as Federated Learning (FL) \cite{mcmahan2017communication}, a distributed paradigm that allows devices to collaboratively train a shared model while keeping data local. By shifting training to end devices, FL places local computation, and consequently energy consumption, at the center of system performance constraints. The energy-aware FL frameworks use techniques such as client selection, pruning, sparsification, and dynamic resource allocation to optimize training energy \cite{li2023anycostfl, yang2020energy, shi2022toward, zeng2023resource}. These methods require an accurate estimation of computation energy, making power modeling accuracy central. In practice, however, most existing state-of-the-art FL frameworks use approximate CPU power models, that estimate CPU power from frequency alone, due to the scarcity of the processor's voltage readings. 

While convenient, such approximations assume linear voltage-frequency scaling and homogeneous cores neglecting key physical parameters and architectural heterogeneity that drive power demand under CPU-intensive workloads. This limitation is worsened on modern mobile SoCs, with multi-cluster CPU architectures \cite{arm_biglittle_whitepaper}, where multiple clusters have distinct performance and energy-efficiency use cases, with appropriate frequency and voltage ranges. Approximate, cluster-agnostic models fail to capture this heterogeneity and misestimate power demand which consequently degrades their optimization strategies.

This paper addresses this apparent gap and presents a reproducible methodology for estimating the dynamic power of heterogeneous multi-cluster ARM-based CPUs by adopting the analytical CMOS-based model. Our approach accounts for cluster heterogeneity and resolves the lack of voltage observability through a rail-to-cluster mapping technique to extract per-cluster supply voltages across operating frequencies. This enables accurate power modeling on commodity mobile devices without additional hardware support or external measurement tools. The key contributions of this paper are:

\begin{itemize}
    \item Estimating CPU dynamic power using the analytical CMOS-based model, accounting for cluster-level heterogeneity in modern ARM-based mobile SoCs.
    \item Extracting per-cluster supply voltage across frequencies, despite the absence of hardware documentation or direct voltage readings.
    \item Demonstrating that analytical modeling achieves errors below 10\%, while approximate models incur errors of up to 959\%.
    \item Highlighting the impact of the power model on the optimization strategy of AnycostFL \cite{li2023anycostfl} as a case study. We showed that the analytical model reached target accuracies (80\% and 90\%) using approximately 1{,}000\,J and 2{,}500\,J respectively, compared to  5{,}000\,J using the approximate model.
\end{itemize}

The rest of this paper is organized as follows: Section \ref{sec:problem_statement} frames the problem and research questions. Section \ref{sec:methodology} and \ref{sec:implementation_details} outline the methodology and implementation. Section \ref{sec:evaluation} validates the approach against approximate methods and demonstrates its impact on AnycostFL. Finally, Section \ref{sec:related_work} reviews relevant literature, and Section \ref{sec:conclusion} concludes the paper.

 \section{Problem Statement and Research Questions}
\label{sec:problem_statement}

In CMOS-based processors, the power draw originates from three main sources: dynamic power due to switching between charging-discharging of capacitance, short-circuit power caused by transient current paths during switching, and leakage power, a static loss consumed even with no switching activity. The integrated power equation is given by \cite{baek2017power}:
\begin{equation}
    P_\text{total} = P_\text{dyn} + P_\text{short\_circuit} + P_\text{leakage}
    \label{eq:cpu_power}
\end{equation}
Dynamic power dominates \cite{burd1995} and depends on frequency $f$, voltage $V$, and effective capacitance $C_\text{eff}=\alpha C$. The analytical model is:
\begin{equation}
P_\text{dyn} = C_\text{eff} V^2 f
\label{eq: exact_dyn_power}
\end{equation}
For full workload, $C_\text{eff}$ is approximately constant, making this model physically grounded \cite{baek2017power,burd1996processor}.

In contrast, many energy-aware FL frameworks \cite{li2023anycostfl, yang2020energy, shi2022toward, zeng2023resource} assume $V \propto f$ and use:
\begin{equation}
    P_\text{dyn} \approx \epsilon f^3
    \label{eq:approx_dyn_power}
\end{equation}
However, even if $\epsilon$ captures voltage scaling and workload effects, it must be re-derived per frequency, limiting generalization and oversimplifying Equation~(\ref{eq: exact_dyn_power}).

A preliminary validation on an Intel Xeon W-2123 \cite{intel_xeon_w2123} (Table~\ref{tab:workstation-proof}) shows that the analytical model achieves errors lower than 1\%, while the approximate model yields errors of $40.6\%$ and $217\%$, highlighting the limitations of the $V \propto f$ assumption, especially for heterogeneous mobile systems. This paper addresses the following research questions:

\begin{table}[!t]
\centering
\small
\caption{Workstation evaluation: Analytical vs. Approximate power models compared to RAPL ground truth.}
\label{tab:workstation-proof}
\resizebox{0.85\columnwidth}{!}{
\begin{tabular}{cccccc}
\toprule
\textbf{Model} & \textbf{Freq [Hz]} & \textbf{Param} & \textbf{$P_\text{dyn}$ [W]} & \textbf{$\hat{P}_\text{dyn}$ [W]} & \textbf{Err [\%]} \\
\midrule
Analytical & $1.2 \times 10^{9}$ & $C_{\text{eff}} = 8.2 \times 10^{-9}$ F & 5.57 & 5.62 & \textbf{0.94} \\
Approximate & $1.2 \times 10^{9}$ & $\epsilon = 1.91 \times 10^{-27}$ & 5.57 & 3.31 & 40.6 \\
\cmidrule{1-6}
Analytical & $3.6 \times 10^{9}$ & $C_{\text{eff}} = 8.2 \times 10^{-9}$ F & 28.21 & 27.95 & \textbf{0.95} \\
Approximate & $3.6 \times 10^{9}$ & $\epsilon = 1.91 \times 10^{-27}$ & 28.21 & 89.3 & 217 \\
\bottomrule
\end{tabular}
}
\end{table}
 
\begin{itemize}
    \item What is the accuracy of the approximate model compared to the analytical CMOS-based model on heterogeneous multi-cluster mobile devices? 
    \item How do approximate power models affect energy estimation and optimization decisions in energy-aware FL frameworks? 
\end{itemize}
We address these by proposing a cluster-aware power estimation methodology (Section \ref{sec:methodology}), evaluating model accuracy on a real-world testbed and investigating implications for AnycostFL (Section~\ref{sec:fl_energy_implications}).
\section{Methodology}
\label{sec:methodology}

Processors based on x86 and ARM technologies differ fundamentally in their architecture and system-level capabilities. While x86 platforms provide direct CPU power access via the RAPL interface \cite{rapl_ref}, ARM devices expose only battery-level power through the fuel gauge \cite{android_device_power}. This together with differences in voltage visibility, core isolation, and thermal behavior necessitate a dedicated methodology for ARM-based mobile devices. Modern ARM SoCs are built with multiple heterogeneous clusters.

A common architecture follows ARM's big.LITTLE technology \cite{arm_biglittle_whitepaper}, where LITTLE cores are destined for energy-efficiency, while big cores handle demanding workloads. A more advanced architecture employs a tri-cluster CPU configuration (e.g., Google Tensor SoCs), with additional ultra-performance cores. Each cluster is a computational unit, with independent voltage and frequency values, and can operate simultaneously or powered off when needed, allowing for per-cluster power measurements. 

\begin{figure}[t]
    \centering
    \includegraphics[width=1\textwidth]{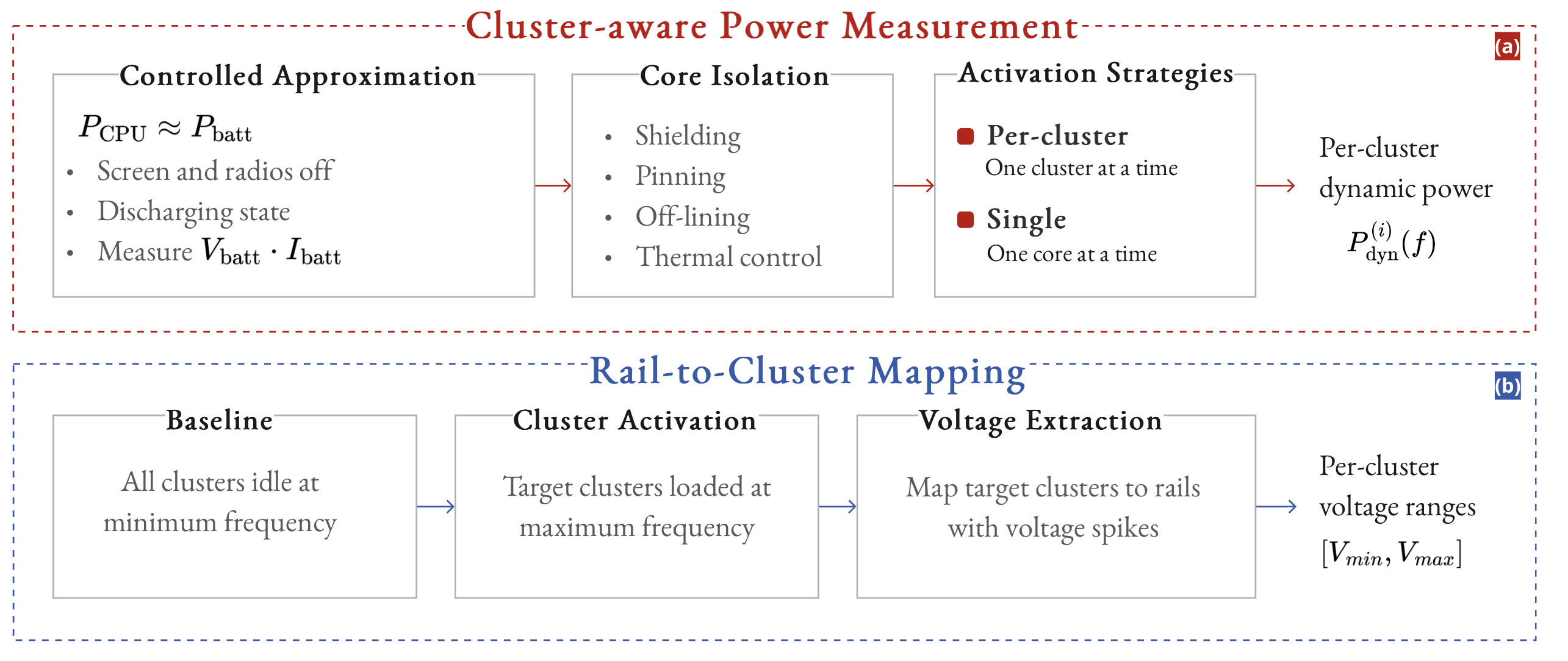}
    \caption{Main pipeline for computing CPU dynamic power on heterogeneous ARM-based mobile devices. \textbf{(a)} Controlled conditions and core isolation techniques applied on two Core/Cluster activation strategies (e.g., Per-cluster and Single) to measure per-cluster dynamic power $P^{(i)}_{\mathrm{dyn}}(f)$ , \textbf{(b)} Rail-to-cluster mapping to extract voltage ranges $(V_{\min},V_{\max})$ per operating-frequency.}
    \label{fig:main_pipeline}
\end{figure}

To the best of our knowledge, this is the first attempt to provide such a fine-grained methodology for modern mobile and IoT devices with heterogeneous multi-cluster CPUs. Figure \ref{fig:main_pipeline} depicts our proposed methodology built on two components: (i) cluster-aware activation-based power measurement strategies to measure dynamic power (see Section \ref{sec:measurement_principles}), and (ii) a rail-to-cluster mapping technique to extract per-cluster supply voltages (see Section \ref{sec:mobile_voltage}), enabling the analysis and validation of the analytical power model in Equation (\ref{eq: exact_dyn_power}). 

\subsection{Battery-to-CPU Power Approximation}
Tools and interfaces for fine-grained CPU power measurements in mobile and IoT devices are not yet standardized \cite{mobile_power_tools_survey} and often rely on external hardware, as recommended by the Android platform for component-level power analysis \cite{android_component_power}. Consequently, many prior studies use external power analyzers such as Monsoon \cite{monsoon_3} and BattOR \cite{battor_1}. While these tools produce high-precision rail-level power measurements, they call for invasive procedures and non-portable physical setup with limited scalability. In light of these constraints, software-based power monitoring represents an accessible and inexpensive alternative for power measurement on mobile phones. 

Battery power reflects the power drawn by the device's components such as CPU, display, radios, DRAM, etc., given by:
\begin{equation}
    P_\text{batt} = P_\text{CPU} + P_\text{disp} + P_\text{radio} + P_\text{misc}
\end{equation}
where \(P_\text{disp}\), $P_\text{radio}$ and $P_\text{misc}$ denote the power demand of the display subsystem, wireless interfaces and other components respectively. According to prior works \cite{carroll2010,pathak2011fine}, the display, connectivity interfaces and CPU dominate the device's overall power draw. By minimizing the contribution of dominant non-CPU components, conducting CPU-bound benchmarks, and assuming remaining components to be static, measured battery variations can be attributed chiefly to CPU activities ($P_\text{CPU} \approx P_\text{batt}$).

Battery power \(P_\text{batt}\) can be expressed in terms of the battery supply voltage and current, both exposed by the fuel gauge:
\begin{equation}
    P_\text{batt} = V_\text{batt} \cdot I_\text{batt}
    \label{eq:battery_power}
\end{equation}
The Power Profiler \cite{boubouh2023power} tool is used to compute average battery power demand using Equation (\ref{eq:battery_power}). While this approach provides device-level power, it does not account for individual clusters within heterogeneous mobile SoCs. Thus, next we introduce our proposed cluster-aware dynamic power estimation approach.

\subsection{Cluster-aware Dynamic Power}
\label{sec:measurement_principles}

\begin{table}[!t]
\centering
\small
\caption{CPU dynamic power measurement protocol at minimum and maximum frequencies.}
\label{tab:dyn_protocol}
\resizebox{0.8\columnwidth}{!}{
\begin{tabular}{llll}
\toprule
\textbf{Phase} & \textbf{Freq/Governor} & \textbf{State} & \textbf{Output} \\
\midrule
\texttt{idle\_min}   & \texttt{powersave} / $f_{\min}$   & Idle & $P_{\text{idle}}(f_{\min})$ \\
\texttt{stress\_min} & \texttt{powersave} / $f_{\min}$   & $\sim$100\% load & $P_{\text{load}}(f_{\min})$ \\
\texttt{idle\_max}   & \texttt{performance} / $f_{\max}$ & Idle & $P_{\text{idle}}(f_{\max})$ \\
\texttt{stress\_max} & \texttt{performance} / $f_{\max}$ & $\sim$100\% load & $P_{\text{load}}(f_{\max})$ \\
\midrule
\multicolumn{4}{l}{\textbf{Dynamic power:} $P_{\text{dyn}}(f)=P_{\text{load}}(f)-P_{\text{idle}}(f)$, $f\in\{f_{\min}, f_{\max}\}$.} \\
\bottomrule
\end{tabular}
}
\end{table}

Dynamic power is defined as the difference between the power drawn by a loaded CPU and an idle baseline. For each considered frequency point $f$, we assume two phases: an \emph{idle} phase, measuring the average power of online but unloaded core(s), denoted \(P_{\text{idle}}(f)\), and a \emph{loaded} phase, where CPU core(s) are fully stressed using \texttt{stress-ng} \cite{stress-ng}, denoted \(P_{\text{load}}(f)\). 

We select two operating points: the minimum frequency under the \emph{powersave} governor and the maximum frequency under the \emph{performance} governor. Table \ref{tab:dyn_protocol} shows four measurement phases. Each phase is run for a sufficient duration and repeated multiple times, from which average values are reported. On multi-cluster CPUs, we record the average battery power while activating a target cluster \({Cl_i}\), and treat it as an estimate of that cluster's dynamic power, $P^{(i)}_{\mathrm{dyn}}(f)$. We propose two strategies for the activation of core(s) to isolate dynamic power.

\subsubsection{Per-cluster Activation}
\label{sec:per_cluster_activation}
This strategy (Algorithm~\ref{alg:percluster}) isolates the dynamic power of a target cluster \({Cl_i}\) by keeping it online while all other clusters are switched off. This controlled setup attributes any power variation to the target cluster. During the loaded phase, all cores in the cluster are driven to full utilization, except for the housekeeping core (e.g., core 0), which is reserved for system tasks. The dynamic power contribution of cluster \({Cl_i}\) is:
\begin{equation}
P_{\mathrm{dyn}}^{(i)}(f) = P_{\mathrm{load}}^{(i)}(f)- P_{\mathrm{idle}}^{(i)}(f).
\end{equation}
The total power demand of the CPU then becomes:
\begin{equation}
    P_{CPU}(f)=\sum_{i=1}^{|Cl|} P_{\mathrm{dyn}}^{(i)}(f) 
    \label{eq:dynamic_CPU_cluster}
\end{equation}

\subsubsection{Single Activation}
\label{sec:single_activation}
The Single activation strategy (Algorithm \ref{alg:single}) captures fine-grained dynamic power contribution of individual cores. Within the target cluster, only one core $k$ and the housekeeping core are kept online at a time. The dynamic power contribution of an individual core \( k \) is:
\begin{equation}
P_{core}^k(f) = \left[P_{\mathrm{load}}^{k}(f) + P_{\mathrm{idle}}^{k_0}(f)\right] - P_{\mathrm{idle}}^{k_0+k}(f),
\end{equation}
where $P_{\mathrm{idle}}^{k_0+k}(f)$ denotes the idle power of both cores, and $P_{\mathrm{load}}^{k}(f)$ is the power of core $k$ when fully loaded. This iterates through each core $k$ in a cluster, deducing its dynamic power as: 
\begin{equation}
P_{\mathrm{dyn}}^{(i)}(f) = \sum_{k \neq k_0} P_{core}^k(f).
\end{equation}

\subsection{Rail-to-cluster Voltage Mapping}
\label{sec:mobile_voltage}

Analytical power modeling requires $(V,f)$ pairs, yet on heterogeneous multi-cluster architectures, the frequency-voltage relationship is not linear nor consistent across clusters. We propose a rail-to-cluster voltage mapping procedure to retrieve per-cluster supply voltage. In modern SoCs, each cluster is powered by a dedicated rail, but these lack public documentation. We reverse-engineered DVFS by monitoring regulator rails exposed by the Linux kernel while activating clusters at different frequencies and workloads. 

For each cluster, a CPU-bound workload is pinned to all its cores while remaining clusters are kept idle. By logging regulator voltage, we track the rails whose voltage increases when the target cluster is activated. Repeating this across all clusters enables mapping each rail to its cluster and identifying associated voltage ranges. Each voltage spike corresponds to the activation of a specific cluster, allowing direct mapping of minimum and maximum supply voltages \((V_{\min}, V_{\max})\). 

\subsection{Power Model Validation}
\label{sec:power_model_validation}

Once dynamic power \(P_\text{dyn}(f)\) is measured supply voltages are obtained the effective capacitance \(C_\text{eff}\) can be derived by:
\begin{equation}
C_\text{eff}(f) = \frac{P_\text{dyn}(f)}{f.V^2} 
\label{eq:c_eff_exact}
\end{equation}
For the approximate model, $\epsilon$ is derived by reversing Equation (\ref{eq:approx_dyn_power}):
\begin{equation}
\epsilon(f) \approx \frac{P_\text{dyn}(f)}{f^3} 
\label{eq:c_eff_approx}
\end{equation}
With the analytical model, \(C_{\text{eff}}\) is expected to remain constant for a given micro-architecture and 100\% workload (e.g., $\alpha=1$), besides measurement noise at minimum and maximum frequency. For the approximate model, the difference between $\epsilon(f_{min})$ and $\epsilon(f_{max})$ is typically significant; therefore, we define a single representative value of $\epsilon$ as the arithmetic mean:
\begin{equation}
    \label{eq:mean_epsilon}
    \epsilon
  = \frac{\epsilon(f_{min}) + \epsilon(f_{max})}{2}.
\end{equation}
With \(C_{\text{eff}}\), \(\epsilon\), operating frequencies and their corresponding supply voltages available, power can be estimated using the analytical and approximate models in Equations (\ref{eq: exact_dyn_power}) and (\ref{eq:approx_dyn_power}). 

 We follow a validation procedure to evaluate accuracy using the relative prediction error:
\begin{equation}
\text{Error} = \frac{\hat{P}_\mathrm{dyn}(f) - P_{\mathrm{dyn}}(f)}{P_{\mathrm{dyn}}(f)} \times 100\%
\end{equation}
Lower prediction error reflects agreement between model predictions and CMOS circuit behavior. We consider errors below 5\% acceptable, as measurements are prone to noise and thermal fluctuations.
\section{Implementation}
\label{sec:implementation_details}

\subsection{System Configuration}

Rail-to-cluster mapping and fine-grained power measurements share common setup steps: frequency and governor configuration, cluster/core isolation, and workload execution. Direct \texttt{sysfs} control is typically restricted by SELinux; therefore, we use the EX Kernel Manager (EXKM) \cite{EXKernelManager} to set per-cluster frequencies and switch between \texttt{powersave} and \texttt{performance} governors. To disable DVFS (Dynamic Voltage and Frequency Scaling), minimum and maximum frequencies are set to identical values before each phase. 

Isolation is achieved via Linux kernel mechanisms: \texttt{cpuset} cgroups shield the \texttt{SYSTEM\_CORE} (core 0) from background tasks; cores are toggled via \path{/sys/devices/system/cpu/cpuX/online}; and workloads are bound to target cores using \texttt{taskset}:
\begin{verbatim}
taskset -c k stress-ng --cpu 1 --timeout T
\end{verbatim}
where \texttt{-c k} pins the \texttt{stress-ng} worker to core \texttt{k} for duration \texttt{T}. All scripts for configuration, isolation, and mapping are available in our public repository \cite{arm_power_repo}. 

\subsection{Power Measurement Setup}

Loaded phases employ \texttt{stress-ng} with the \texttt{--cpu-method all} option to ensure workload-independent stress. Experiments are executed via the \texttt{Termux} terminal emulator \cite{termux}. Battery-level power is logged using Power Profiler \cite{boubouh2023power}, which samples $V_{\text{batt}}$, $I_{\text{batt}}$, frequency, utilization, and temperature every 0.5\,s.

To mitigate thermal throttling bias, we enforce a target CPU temperature of 30\,$^\circ$C \cite{luo2008system}. If the temperature deviates, we employ dynamic warming (multi-core stress) or cooling (core off-lining and idling). To ensure stability, each 10-minute phase is repeated 5 times. An idle-before-load order is maintained, with consistent thermal management between runs. Results are reported as the mean across runs, with variability indicated by min--max ranges or standard deviation.

\subsection{Activation Strategies}
\label{sec:activation_algorithms}

We introduce two activation strategies to isolate cluster-level dynamic power.

\textbf{Per-cluster Activation:} Algorithm~\ref{alg:percluster} performs the per-cluster activation procedure by off-lining non-target clusters, measuring power in idle and loaded states, then extracting the per-cluster dynamic power as their difference \cite{arm_percluster_script}.

\textbf{Single Activation:} Algorithm~\ref{alg:single} isolates per-core dynamic power by measuring one core at a time while keeping the SYSTEM\_CORE $k_0$ online for OS activities. It alternates idle and loaded phases for each target core and computes the cluster dynamic power as the sum of per-core contributions ~\cite{arm_single_script}.

\begin{algorithm}[h]
\caption{Per-cluster Activation}
\label{alg:percluster}
\begin{algorithmic}[1]
\Require Target cluster $Cl_i$, frequency $f$, SYSTEM\_CORE $k_0$
\Ensure $P^{(i)}_{\mathrm{dyn}}(f)$
\State Offline all clusters except $Cl_i$; keep all cores in $Cl_i$ online
\State Pin $Cl_i$ to frequency $f$; shield system tasks on $k_0$
\State $P^{(i)}_{\mathrm{idle}}(f) \gets \textsc{MeasureAvgPower}()$
\State Pin \texttt{stress-ng} to all cores $k \in Cl_i$ where $k \neq k_0$
\State $P^{(i)}_{\mathrm{load}}(f) \gets \textsc{MeasureAvgPower}()$
\State $P^{(i)}_{\mathrm{dyn}}(f) \gets P^{(i)}_{\mathrm{load}}(f) - P^{(i)}_{\mathrm{idle}}(f)$; \Return $P^{(i)}_{\mathrm{dyn}}(f)$
\end{algorithmic}
\end{algorithm}

\begin{algorithm}[h]
\caption{Single Activation}
\label{alg:single}
\begin{algorithmic}[1]
\Require Target cluster $Cl_i$, frequency $f$, SYSTEM\_CORE $k_0$
\Ensure $P_{\mathrm{dyn}}^{(i)}(f)$, $\{P_{\mathrm{core}}^{k}(f)\}$
\State Offline clusters except $Cl_i$; pin $f$; keep only $k_0$ online and shielded
\State $P_{\mathrm{dyn}}^{(i)}(f) \gets 0$
\ForAll{$k \in Cl_i, k \neq k_0$}
    \State Bring core $k$ online; $P_{\mathrm{idle}}^{k_0+k}(f) \gets \textsc{MeasureAvgPower}()$
    \State Pin \texttt{stress-ng} to core $k$; $P_{\mathrm{load}}^{k}(f) \gets \textsc{MeasureAvgPower}()$
    \State $P_{\mathrm{core}}^{k}(f) \gets \left[P_{\mathrm{load}}^{k}(f) + P_{\mathrm{idle}}^{k_0}(f)\right]- P_{\mathrm{idle}}^{k_0+k}(f)$
    \State $P_{\mathrm{dyn}}^{(i)}(f) \gets P_{\mathrm{dyn}}^{(i)}(f) + P_{\mathrm{core}}^{k}(f)$; Offline core $k$
\EndFor
\State \Return $P_{\mathrm{dyn}}^{(i)}(f)$
\end{algorithmic}
\end{algorithm}
\section{Evaluation}
This section introduces the hardware specifications of our experimental testbed, then addresses the research questions in Section \ref{sec:problem_statement} by evaluating the cluster-aware power measurement methodology presented in Section \ref{sec:methodology}.
\label{sec:evaluation}

\subsection{Hardware Configuration} 
\label{sec:hardware_phones}

To analyze cluster-specific dynamic power on ARM-based mobile platforms, two Android devices hosted the experiments, and whose hardware and software configuration are summarized in Table \ref{table:phones_specs}. The Google Pixel 8 Pro integrates a Google Tensor~G3 SoC with a heterogeneous tri-cluster CPU (LITTLE, big and Prime), representing upper-tier mobile platforms.
The Samsung A16, powered by Mediatek Helio G99 SoC, adopts a big.LITTLE CPU design and is representative of lower-cost devices with more modest computational capabilities. 

Both devices reflect the heterogeneity of modern commodity devices in terms of performance, architecture and energy characteristics. We obtained the minimum and maximum operating voltages of each CPU cluster at corresponding operating frequencies using the rail-to-cluster mapping presented in Section \ref{sec:mobile_voltage}. Table \ref{tab:cluster-voltage-table} presents the voltage ranges for both Google Pixel 8 Pro and Samsung A16. 

\begin{table}[t!]
\centering
\caption{Hardware specifications of the Android devices used in our experiments.}
\label{table:phones_specs}
\resizebox{\columnwidth}{!}{
\begin{tabular}{lcccc}
\toprule
\textbf{Device} & \textbf{SoC} & \textbf{Clusters} & \textbf{RAM} & \textbf{Operating System} \\
\midrule
Google Pixel 8 Pro 
& Google Tensor G3 
& LITTLE + big + Prime 
& 12~GB 
& Android 14 \\
Samsung A16 
& MediaTek Helio G99 
& LITTLE + big 
& 8~GB 
& Android 14 \\
\bottomrule
\end{tabular}
}
\end{table}

\begin{table}[t!]
\centering
\caption{Measured per-cluster operating ranges (frequency and voltage) for two mobile SoCs used in our experiments.}
\label{tab:cluster-voltage-table}
\resizebox{\columnwidth}{!}{
\begin{tabular}{lcccccc}
\toprule
\textbf{Device} & \textbf{Cluster} & \textbf{Cores} & \textbf{$f_{\min}$ [Hz]} & \textbf{$f_{\max}$ [Hz]} & \textbf{$V_{\min}$ [V]} & \textbf{$V_{\max}$ [V]} \\
\midrule

\multirow{3}{*}{Google Pixel 8 Pro (Tensor G3)}
& LITTLE & 4 & $3.24 \times 10^{8}$ & $1.70 \times 10^{9}$ & 0.56 & 0.85 \\
& big    & 4 & $4.02 \times 10^{8}$ & $2.37 \times 10^{9}$ & 0.55 & 1.13 \\
& Prime  & 1 & $5.00 \times 10^{8}$ & $2.91 \times 10^{9}$ & 0.53 & 1.20 \\
\midrule

\multirow{2}{*}{Samsung A16 (MediaTek Helio G99)}
& LITTLE & 6 & $5.00 \times 10^{8}$ & $2.00 \times 10^{9}$ & 0.55 & 0.81 \\
& big    & 2 & $7.25 \times 10^{8}$ & $2.20 \times 10^{9}$ & 0.55 & 0.76 \\
\bottomrule
\end{tabular}
}
\end{table}

\subsection{Experimental Results}
\label{sec:experimental_results}

\subsubsection{Activation Algorithm Validation}
\label{sec:activation_validation}

\begin{table*}[!t]
\centering
\caption{Analytical model accuracy across activation strategies. $P_{dyn}$ is the measured dynamic power ($\pm$ standard deviation). Predicted power $\hat{P}$ is computed using the averaged $C_{\mathrm{eff}}$. Bolding indicates the strategy with the lowest absolute error for each cluster.}
\label{tab:analytical_accuracy_final}
\resizebox{\textwidth}{!}{
\begin{tabular}{lcl|ccc|ccc}
\toprule
\textbf{Device} & \textbf{Activation Strategy} & \textbf{Cluster} & 
\multicolumn{3}{c|}{\textbf{Min frequency}} & 
\multicolumn{3}{c}{\textbf{Max frequency}} \\

 &  &  & 
$P_{dyn}$ [W] & $\hat{P}$ [W] & Err [\%] & 
$P_{dyn}$ [W] & $\hat{P}$ [W] & Err [\%] \\

\midrule

\multirow{4}{*}{Samsung A16} 
& \multirow{2}{*}{Per-cluster} & LITTLE 
& 0.182 {\scriptsize $\pm$0.087} & 0.099 & 8.5 
& 0.549 {\scriptsize $\pm$0.074} & 0.825 & -7.3 \\
& & big 
& 0.189 {\scriptsize $\pm$0.062} & 0.198 & 4.8 
& 0.806 {\scriptsize $\pm$0.042} & 0.787 & -4.4 \\
\cmidrule(lr){2-9}
& \multirow{2}{*}{Single} & LITTLE 
& 0.100 {\scriptsize $\pm$0.045} & 0.102 & \textbf{1.6} 
& 0.859 {\scriptsize $\pm$0.143} & 0.846 & \textbf{-1.5} \\
& & big 
& 0.206 {\scriptsize $\pm$0.037} & 0.211 & \textbf{2.5} 
& 0.862 {\scriptsize $\pm$0.081} & 0.841 & \textbf{-2.4} \\

\midrule

\multirow{6}{*}{Pixel 8 Pro} 
& \multirow{3}{*}{Per-cluster} 
& LITTLE 
& 0.146 {\scriptsize $\pm$0.041} & 0.135 & -8.0 
& 0.995 {\scriptsize $\pm$0.097} & 1.090 & 9.6 \\

& & big 
& 0.142 {\scriptsize $\pm$0.095} & 0.157 & 10.2 
& 4.267 {\scriptsize $\pm$0.101} & 3.910 & -8.5 \\

& & Prime 
& 0.100 {\scriptsize $\pm$0.065} & 0.102 & \textbf{2.0} 
& 3.114 {\scriptsize $\pm$0.063} & 3.050 & \textbf{-2.0} \\

\cmidrule(lr){2-9}

& \multirow{3}{*}{Single} 
& LITTLE 
& 0.142 {\scriptsize $\pm$0.070} & 0.136 & \textbf{-3.9} 
& 1.056 {\scriptsize $\pm$0.167} & 1.100 & \textbf{4.3} \\

& & big 
& 0.199 {\scriptsize $\pm$0.107} & 0.193 & \textbf{-3.1} 
& 4.639 {\scriptsize $\pm$0.153} & 4.790 & \textbf{3.3} \\

& & Prime 
& 0.100 {\scriptsize $\pm$0.021} & 0.103 & 3.1 
& 3.178 {\scriptsize $\pm$0.092} & 3.080 & -2.9 \\

\bottomrule
\end{tabular}
}
\end{table*}

In Section \ref{sec:measurement_principles}, we proposed two activation strategies against ground-truth measurements at the lowest and highest supported frequencies. Overall, both strategies yield reasonable estimates with low relative errors, while the Single strategy outperforms the Per-cluster strategy and achieves errors below 5\% across all clusters and frequencies. For the Samsung A16, the Single strategy achieves an error of only 1.5\% for the LITTLE cluster at maximum frequency, while the Per-cluster strategy reaches errors up to 8.5\%. 

Due to the high granularity of the Single strategy, it requires long per-core configurations. Nevertheless, it is more robust against inter-cluster power interference and thermal throttling. While Per-cluster strategy shows larger errors, between 7\% and 10\%, likely due to noise caused by simultaneous cluster activation and shared voltage noise on commodity heterogeneous SoCs. Since Single outperforms Per-cluster by achieving lower errors and better repeatability, we adopt it for the remainder of this paper and subsequent experiments.  

\subsubsection{Analytical vs. Approximate Model Accuracy}
\label{sec:accuracy_models}

\begin{figure}[t]
    \centering
    \includegraphics[width=1\textwidth]{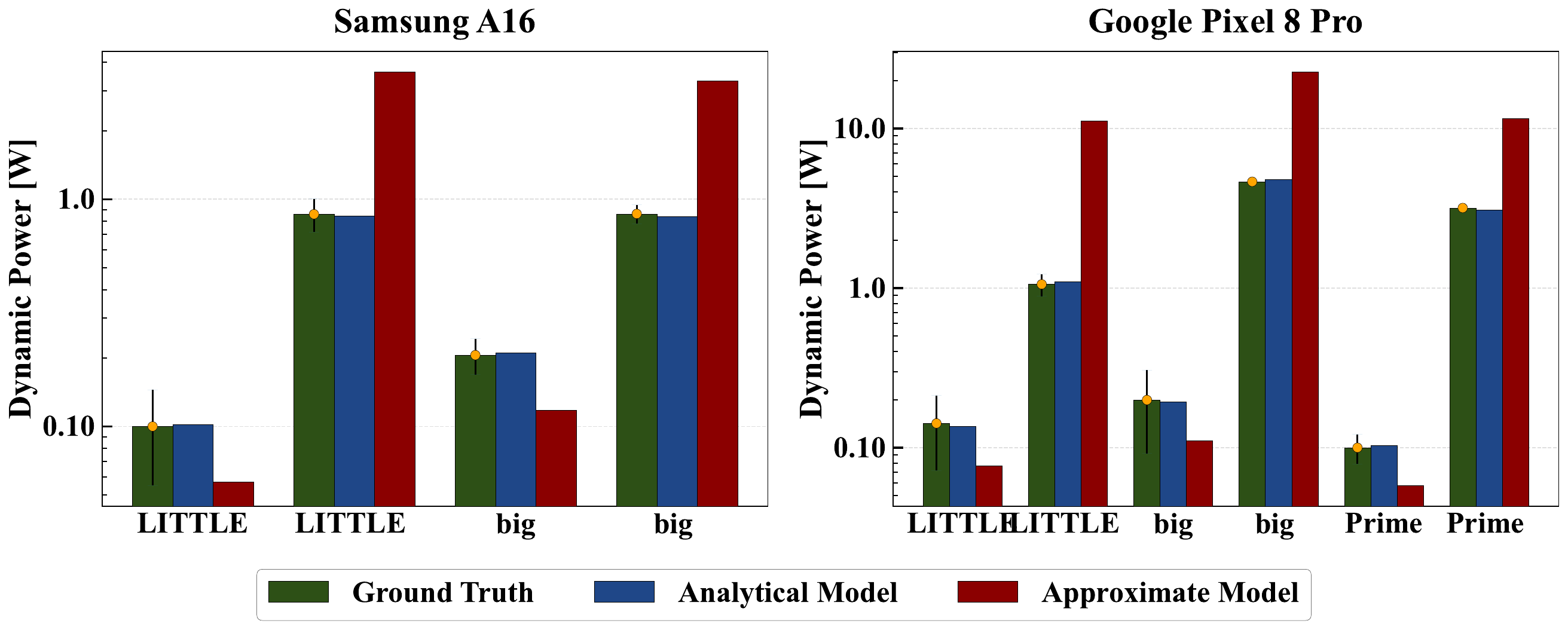}
    \caption{Dynamic power prediction comparison across analytical and approximate models on Samsung A16 and Google Pixel 8 Pro devices. Bars show mean predicted across 5 runs: ground truth power, analytical power model and approximate power model. The analytical model closely tracks ground truth values across all clusters and frequency points, while the approximate model deviates from measured power by factors up to 2-5x.}
    \label{fig:absolute_error}
\end{figure}

\begin{table*}[!t]
\centering
\small
\caption{Reported average values of ground truth ($P$) versus predicted power ($\hat{P}$) using analytical and approximate power models (Single Strategy). Relative error values represent mean $\pm$ std across runs.}
\label{tab:absolute_error_single}
\resizebox{\textwidth}{!}{
\begin{tabular}{llc|ccc|ccc}
\toprule
\textbf{Device} & \textbf{Cluster} & \textbf{Freq. [Hz]} & \textbf{$P$ [W]} & \multicolumn{2}{c|}{\textbf{Analytical}} & \multicolumn{3}{c}{\textbf{Approximate}} \\
\cmidrule(lr){5-6} \cmidrule(lr){7-9}
& & & & $\hat{P}$ [W] & Err [\%] & $\hat{P}$ [W] & Err [\%] \\

\midrule

\multirow{4}{*}{Samsung A16} 
& \multirow{2}{*}{LITTLE} 
& $5.00 \times 10^{8}$ & 0.100 {\scriptsize $\pm$0.045} & 0.102 & \textbf{1.6} & 0.057 & -43.3 \\
& & $2.00 \times 10^{9}$ & 0.859 {\scriptsize $\pm$0.143} & 0.846 & \textbf{-1.5} & 3.630 & 322.0 \\
\cmidrule(lr){2-8}
& \multirow{2}{*}{big} 
& $7.25 \times 10^{8}$ & 0.206 {\scriptsize $\pm$0.037} & 0.211 & \textbf{2.5} & 0.118 & -42.5 \\
& & $2.20 \times 10^{9}$ & 0.862 {\scriptsize $\pm$0.081} & 0.841 & \textbf{-2.4} & 3.310 & 284.0 \\

\midrule

\multirow{6}{*}{Pixel 8 Pro} 
& \multirow{2}{*}{LITTLE} 
& $3.24 \times 10^{8}$ & 0.142 {\scriptsize $\pm$0.070} & 0.136 & \textbf{-3.9} & 0.077 & -45.8 \\
& & $1.70 \times 10^{9}$ & 1.056 {\scriptsize $\pm$0.167} & 1.100 & \textbf{4.3} & 11.200 & 959.0 \\
\cmidrule(lr){2-8}
& \multirow{2}{*}{big} 
& $7.25 \times 10^{8}$ & 0.199 {\scriptsize $\pm$0.107} & 0.193 & \textbf{-3.1} & 0.111 & -44.3 \\
& & $2.20 \times 10^{9}$ & 4.639 {\scriptsize $\pm$0.153} & 4.790 & \textbf{3.3} & 22.600 & 388.0 \\
\cmidrule(lr){2-8}
& \multirow{2}{*}{Prime} 
& $5.00 \times 10^{8}$ & 0.100 {\scriptsize $\pm$0.021} & 0.103 & \textbf{3.1} & 0.058 & -42.0 \\
& & $2.91 \times 10^{9}$ & 3.178 {\scriptsize $\pm$0.092} & 3.080 & \textbf{-2.9} & 11.500 & 262.0 \\

\bottomrule
\end{tabular}
}
\end{table*}

Our first research question, presented in Section \ref{sec:problem_statement}, relates to the accuracy of the analytical model of Equation (\ref{eq: exact_dyn_power}) in contrast to the approximate one of Equation (\ref{eq:approx_dyn_power}). To compare, we use the same prior validation methodology explained in Section \ref{sec:power_model_validation}. Figure \ref{fig:absolute_error} clearly brings out the difference among both models at estimating power by comparing their estimations to ground truth measured power across all clusters and operating frequencies. 

The analytical model shows excellent power prediction behavior, agreeing with the measured values, maintaining prediction errors below 5\%. In contrast, the approximate model deviates significantly from measured power, diverging by a scale of 3 to 10x. These poor predictions are reflected by high relative errors reaching 322.0\% on Samsung A16 and 959.0\% on the Google Pixel 8 Pro at maximum frequencies (See Table \ref{tab:absolute_error_single}). These results demonstrate the unreliability of the approximate model for power estimation compared to the analytical CMOS-based model.

\subsubsection{Implications for Energy-Aware Federated Learning}
\label{sec:fl_energy_implications}
\begin{figure*}[!t]
    \centering
    \includegraphics[width=1\linewidth]{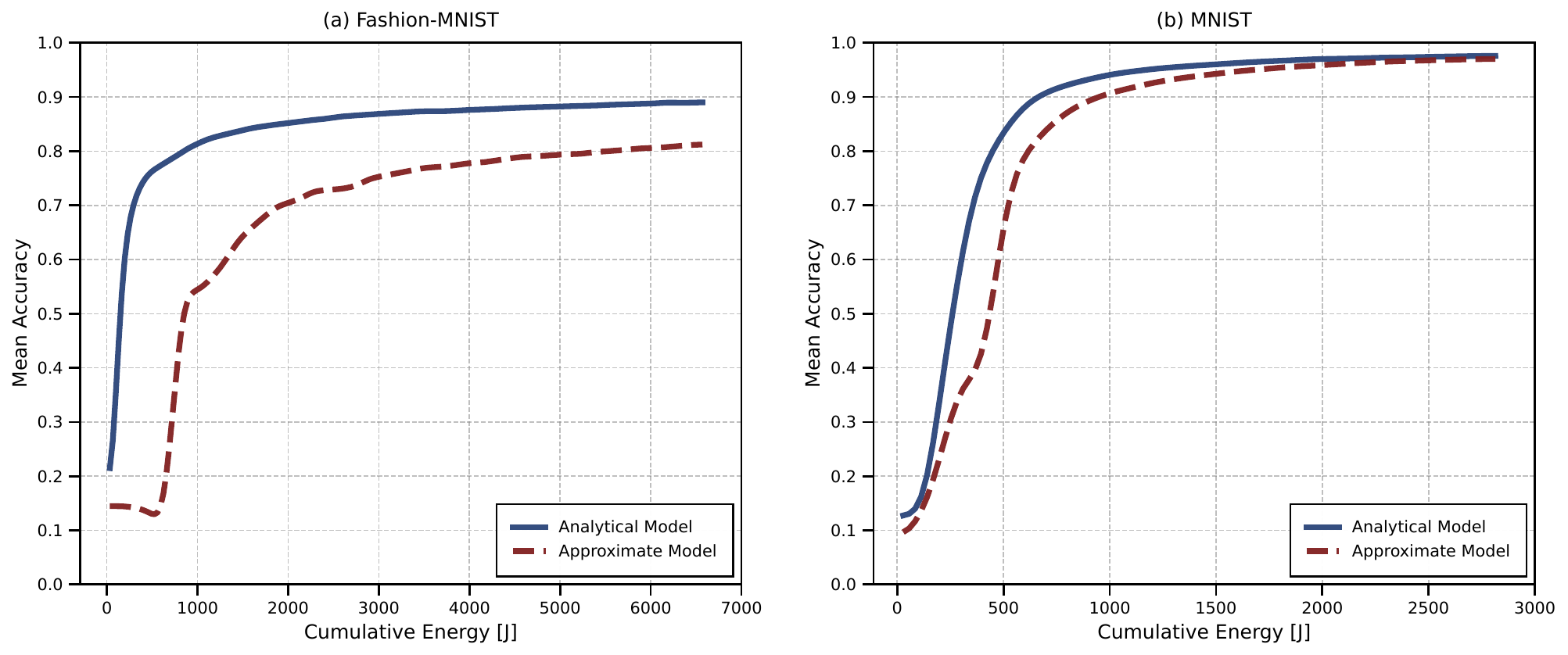}
    \caption{Cumulative computation energy vs. Accuracy of AnycostFL on (a) Fashion-MNIST and (b) MNIST using the analytical and the approximate model. For a given target accuracy, the analytical model showcases lower energy consumption compared to the approximate one, leading to suboptimal optimization decisions that affect model's accuracy.}
    \label{fig:energy_results}
\end{figure*}

Figure \ref{fig:energy_results} reports accuracy versus energy tradeoffs of the AnycostFL framework, trained on Fashion-MNIST and MNIST datasets, when modeling computation energy using the analytical CMOS-based model in Equation (\ref{eq: exact_dyn_power}) or with the approximate formulation commonly used in energy-aware FL frameworks in Equation (\ref{eq:approx_dyn_power}). This figure highlights the paper’s main takeaway: the choice of power model shapes the optimization strategy and subsequent decisions of energy-aware FL (e.g., energy budgets, shrinking factors, resource allocation), and therefore impacts the attainable accuracy under a fixed energy budget. Technical details about the AnycostFL framework and its computation energy as well as the integration of analytical and approximate model are provided in Appendix~\ref{app:anycostfl_details}. 

Across both datasets, and for a target accuracy, the analytical model consumes less computation energy compared to the approximate model. The gap in terms of energy consumption is especially apparent on Fashion-MNIST, where the analytical model reaches 80\% accuracy at roughly 1000J, whereas the approximate model requires more training rounds, consuming above 5000J to attain the same accuracy. On the easier MNIST dataset, both models show similar behavior and converge closely, with the analytical model still outperforming in terms of accuracy with less energy consumption. The same pattern is repeated with the analytical model consuming 600J to achieve a target accuracy of 90\% while the approximate model consumes 1000J to attain the same accuracy. 

\subsection{Discussion}

Figure \ref{fig:energy_results} showcases the divergence between the analytical and approximate models in terms of estimating energy. Since the approximate model does not account for per-cluster voltage domains and effective capacitance, it tends to over-estimate this quantity. As explained in Appendix ~\ref{app:anycostfl_details}, the shrinking factor $\alpha_{t,i}$ is estimated based on the computation energy $E^{\max}_{t,i}$. If over-estimated, the feasible range of $\phi_{t,i}$ becomes tighter, pushing $\alpha_{t,i}$ towards smaller values and training overly small models, leading to slower convergence. We call this phenomenon \emph{over-shrinking}. The analytical model, within $5\%$ from the ground truth, avoids this issue and reaches the same target accuracy with substantially less energy. 

This phenomenon is not specific to AnycostFL, instead, it is prominent on any FL framework basing its energy-aware decisions on computation energy constraints. The methodology is also relevant beyond FL, and can be extended to edge computing frameworks such as on-device inference scheduling, DVFS-aware task placement, etc. With the same pattern appearing on an upper-tier and a lower-cost mobile device (e.g., Google Pixel 8 Pro and Samsung A16), we conclude that ignoring per-cluster voltage domains and effective capacitance is a direct cause of this sensitivity, instead of attributing it to a SoC artifact. 

A natural concern arises when considering the scale of an FL deployment, where many heterogeneous devices are included in the training process. In practice, the characterization cost is correlated with performing the rail-to-cluster mapping and extracting effective capacitance only once. Moreover, extracting parameters needs to be performed once per-SoC, and then re-used for any devices built on it. For a non characterized device, a hybrid approach is possible: apply the analytical model when parameters are ready, and fall back to the approximate model otherwise. This enables potential improvements of power estimation accuracy, without requiring full upfront characterization.

\section{Related Work}
\label{sec:related_work}

Power consumption of processors was firstly analyzed by Chandrakasan et al.~\cite{chandrakasan1992}, who established dynamic power as in Equation (\ref{eq: exact_dyn_power}). Burd and Brodersen~\cite{burd1995} presented an energy-saving framework through supply voltage reduction, demonstrating the high impact of voltage. However, these works only targeted homogeneous cores with uniform voltage domains.

Energy efficiency has also been addressed at the architectural level. Kumar et al.~\cite{kumar2003} demonstrated significant energy savings (15-40\%) by assigning different power-performance traits to cores. This design was later adopted by ARM in big.LITTLE~\cite{greenhalgh2011biglittle} and DynamIQ tri-cluster architectures~\cite{navada2019}. Under these technologies, heterogeneous clusters operate under distinct voltage-frequency points and lack methods to extract per-cluster supply voltage.

Carroll and Heiser~\cite{carroll2010} conducted an early smartphone power study using external hardware, but considered homogeneous cores. Walker et al.~\cite{walker2017} proposed a real-time model based on performance monitoring counters (PMCs) that is cluster-agnostic and data-driven. A CMOS-based model was presented in Baek et al. \cite{baek2015}, which relies on invasive hardware instrumentation and doesn't account for per-cluster voltage and capacitance variations.

In addition to hardware-based measurement techniques, several studies proposed software-based power modeling techniques for power measurements. Pathak et al.~\cite{pathak2011fine} used system activity statistics to build a fine-grained energy accounting framework tested on Android and Windows Mobile. A similar approach named PowerTutor was proposed by Dong and Zhong~\cite{dong2011self} to estimate component-level energy consumption using information from the operating system and the battery. These approaches were able to estimate power without external hardware, yet their prime focus is on system level energy consumption instead of analytical CPU power modeling. Moreover, these studies don't account for heterogeneous multi-cluster SoCs present in modern mobile processors, or consider cluster-level power parameters such as supply voltage and effective switching capacitance, both required to apply the CMOS-based analytical model.

Despite extensive work on CPU power modeling, existing analytical formulations remain difficult to apply on commodity smartphones, mainly due to the absence of per-cluster voltage observability. Consequently, most prior studies either rely on external hardware instrumentation or employ data-driven approximations that remain unaware of heterogeneous cluster-level operating points. To the best of our knowledge, no prior work proposes a reproducible methodology enabling fine-grained analytical modeling of dynamic CPU power demand on real multi-cluster mobile devices.

In this work, we address these limitations by proposing the first comprehensive methodology to enable fine-grained analytical CPU power modeling on commodity heterogeneous mobile devices. Our approach introduces a rail-to-cluster mapping mechanism to retrieve per-cluster supply voltages and presents multiple activation strategies to capture dynamic CPU demand at different granularities. We validate the methodology across several Android smartphones without requiring external measurement equipment or hardware modifications.

\section{Conclusion and Future Work}
\label{sec:conclusion}

This work questions the prevalent reliance on approximate CPU power models in edge and mobile computing and highlights their shortcomings on modern heterogeneous ARM-based platforms. To address this gap, we present a reproducible methodology for estimating dynamic CPU power using the analytical CMOS-based modeling approach. Our methodology overcomes the challenge of unavailable voltage measurements by employing cluster-aware activation and a reverse-engineered mapping from rail to cluster voltages, enabling accurate power estimation without direct voltage observability. 

Our evaluations on two heterogeneous off-the-shelf Android devices showed that the analytical model produces stable, physically meaningful parameters and accurate power estimates, whereas the approximate models used by state-of-the-art energy-aware Federated Learning (FL) frameworks exhibit large errors and poor generalization. When integrated with AnycostFL, a state-of-the-art energy-aware FL framework, we showed that inaccurate approximate models lead to suboptimal energy allocation and poorer accuracy--energy trade-offs, while the analytical model enables more effective use of each device's energy budget.

This work faces several limitations. With our methodology focusing on dynamic CPU power, other contributors to total energy footprint of an FL workload on mobile devices remain unexplored. In practice, the CPU power contains additional power terms (See Equation \ref{eq:cpu_power}), that can have significant impact on total power demand under realistic workloads. As part of our measurement protocol, we include steps to mitigate thermal throttling effects. While useful during the parameter extraction process, long-running FL deployments may experience more intense throttling regimes, altering the power-performance relationship altogether. Finally, the $P_{CPU} \approx \Delta P_{batt}$ remains an under-validated assumption, with the power supply of the battery feeding into other non-CPU components (e.g., DRAM, voltage regulators and SoC interconnect). 

Inspired by these limitations, our future work focuses on extending the analytical model to capture the CPU leakage term, enabling thermal-aware CPU power modeling. We further plan to extend our testbed to include a broader set of SoCs, with ML inference workloads reflecting real-world scenarios, to assess the accuracy of the proposed model under realistic thermal regimes. Overall, this work lays a practical foundation for accurate, portable, and energy-optimal learning on heterogeneous mobile and edge devices.

%
%


\bibliographystyle{plainnat}
\bibliography{references}
\newpage

\appendix

\section{x86-based Workstations}
\label{sec:workstation}

Whether deployed as commodity desktop devices or as servers in a data center environment, most workstations rely on x86-based processors. Our goal in this section is therefore to present a robust methodology for reliably capturing power measurements on such devices, and access the supply voltages readable through model-specific registers.

\subsection{Hardware and Software Details}
In this study, a preliminary validation experiment was conducted on a x86-based Intel workstation, while our proposed methodology was tested on ARM-based mobile devices. The hardware and software specifications for both types of devices are details in what follows. We conducted our preliminary validation experiment on a workstation equipped with a CPU Intel Xeon W-2123 containing 4 physical cores, with one hardware thread each and a single-socket x86 architecture with RAPL (Running Average Power Limit) support. The processor frequencies range from 1.2 GHz to 3.6 GHz. The system runs Ubuntu 24.04.3, a Linux distribution released on 2024. By extracting the VID at both operating points (e.g. minimum and maximum frequency) following the steps in section \ref{sec:ws_voltage}, we deduce the corresponding CPU voltage values of the x86 workstation presented in Table \ref{table:workstation_voltage}. 

\begin{table}[h!]
\centering
\caption{CPU voltage at fixed operating frequencies on the Intel Xeon W\textendash2123 workstation.}
\label{table:workstation_voltage}
\begin{tabular}{lccc}
\toprule
Corner & Frequency [GHz] & VID & $V_{\mathrm{CPU}}$ [V] \\
\midrule
Min  & 1.20 & 6193 & 0.756 \\
Max  & 3.60 & 7971 & 0.973 \\
\bottomrule
\end{tabular}
\end{table}

\subsection{Power Measurement Protocol}
\label{sec:ws_power_measurment}

On the workstation, the open-source software package \texttt{powerstat} \cite{powerstat} is used to extract power demand readings on Linux operating systems, and retrieve additional system statistics such as CPU usage, as well as frequency and temperature values \cite{becker2017software}. To minimize uncontrolled CPU tasks and retrieve a clean and reliable baseline power measurements, a lightweight keyboard-centric window manager named \texttt{i3} \cite{i3wm} is used instead of \texttt{GNOME} \cite{gnome}, the default desktop environment of Ubuntu. 

To reduce measurement noise and instability, \emph{CPU shielding} and \emph{CPU pinning} mechanisms were implemented. Shielding prevents operating system routines and background tasks from running on experimental shielded cores by confining them to a dedicated \texttt{SYSTEM\_CORE}, while pinning fixes the synthetic workload (e.g., \texttt{stress-ng}) on the set of assigned cores. This pinning-and-shielding setup yields more reliable and coherent power measurements, and ensure that power variation is a result of workload rather than noisy OS tasks. 

\subsection{Core Voltage Retrieval}
\label{sec:ws_voltage}

In early x86 platforms (mid-1990s through 2012), CPU Vcore was traditionally accessible through external hardware monitoring chips (e.g., Winbond/Nuvoton W83627, ITE IT87xx) \cite{winbond_w83627_datasheet}. In modern processors, the CPU supply voltage is not exposed through standard OS interfaces such as \texttt{lm\_sensors} or \texttt{hwmonitor} which report this information as unavailable. 

\paragraph{Intel Processors} According to Intel® 64 and IA-32 architectures software developer's manual (vol. 4, MSR reference) \cite{intel_sdm_vol4}, the CPU voltage is encoded as a VID field in an internal voltage–frequency table and can be retrieved from the \texttt{IA32\_PERF\_STATUS (MSR~0x198) } register, precisely in the slice 47:32, scaled by a factor of 1/8192 \cite{piersma2025reverse,intelcommunity_freqscale}. The CPU voltage can be computed as: \begin{equation}
    V_{\text{CPU}} = MSR\_PERF\_STATUS[47\!:\!32] \times 2^{-13}
    \label{VID}
\end{equation}
where \texttt{MSR\_PERF\_STATUS[47:32]} is the VID entry in the register \texttt{IA32\_PERF\_STATUS}, and the factor \(2^{-13}\) converts this fixed-point VID code into a voltage value in volts.

\paragraph{AMD processors}

In AMD processors, voltage information is also encoded as a VID, but exposed differently through P-states in \texttt{MSR\_PSTATE\_n} registers (typically \texttt{MSR 0xC0010064} through \texttt{0xC001006B}). Each P-state entry contains a VID field containing the AMD's Serial Voltage Identification encoding. The supply voltage is found through the conversion of the SVI2 formula defined in the AMD BIOs \cite{amd_bkdg_svi2} and the Kernel Developer's Guide (BKDG) \cite{amd_zen_bkdg} as follows: 
\begin{equation}
V_{\mathrm{CPU}} = V_{\mathrm{offset}} - k \cdot \mathrm{VID}
\label{eq:amd_vid}
\end{equation}
The generation-dependent constants in Equation (\ref{eq:amd_vid}) are available through the BKDG documentation.

\section{AnycostFL: Energy-Aware FL Case Study}
\label{app:anycostfl_details}

AnycostFL \cite{li2023anycostfl} is a state-of-the-art on-demand Federated Learning framework that reduces computation energy using a model-shrinking strategy and allows to train the same network at different widths according to a shrinking factor $\alpha$. In other words, during each local training round, each peer selects its model size according to its resource constraints. This approach balances accuracy and energy among peers with heterogeneous performance ranges and constitutes a solid energy-aware FL framework on mobile and edge devices. 

To illustrate the gap between analytical and approximate models and their impact on energy-aware Federated Learning (FL), we adopt the AnycostFL framework's computation model for time and energy on our experimental devices (see Section \ref{sec:hardware_phones}). For comparison, we compute the energy model twice: once with the widely adopted approximation model of Equation (\ref{eq:approx_dyn_power}) by the energy-aware FL frameworks, and once with our proposal of using the analytical model of Equation (\ref{eq: exact_dyn_power}).

More precisely, for each model, we measure computational energy consumed by peer $i$ at round $t$, given a CPU frequency $f_{t,i}$. Using the analytical dynamic-power model in Equation (\ref{eq: exact_dyn_power}), the computation energy is given by:
\begin{equation}
E^{\mathrm{cmp}}_{t,i} = C_{\mathrm{eff}}^{(i)}\, V_{i}(f_{t,i})^{2}\, \mathcal{W}_{t,i},
\label{eq:fl_energy_std}
\end{equation}
where $C_{\mathrm{eff}}^{(i)}$ is the effective switching capacitance of peer $i$ and $V_i(f_{t,i})$ is the supply voltage corresponding to $f_{t,i}$.
Similarly to Equation (\ref{eq:fl_energy_std}), the approximate energy is modeled as:
\begin{equation}
E^{\mathrm{cmp}}_{t,i} = \epsilon_i\, f_{t,i}^{2}\,\mathcal{W}_{t,i},
\label{eq:fl_energy}
\end{equation}
where $\epsilon_i$ is a device-dependent hardware coefficient. The local compute workload is defined as
\begin{equation}
\mathcal{W}_{t,i} \triangleq \tau\,|D_i|\,\alpha_{t,i}\,W_{\text{sample}}
\label{eq:local_work}
\end{equation}
with $\tau$ the number of local epochs per round, $|D_i|$ the size of peer $i$'s dataset, $\alpha_{t,i}\in[0,1]$ the fraction of samples processed at round $t$, and $W_{\text{sample}}$ the average number of CPU cycles required to process one training sample on the target platform.

\end{document}